---------------------------------------------------------------
\documentstyle[12pt,amssymbols]{article}
\newcommand{\prd}{Phys. Rev. \underline}
\newcommand{\pl}{Phys. Lett. \underline}

\newcommand{\np}{Nucl. Phys. \underline}

\newcommand{\be}{\begin{equation}}
\newcommand{\ee}{\end{equation}}
\newcommand{\bea}{\begin{eqnarray}}
\newcommand{\eea}{\end{eqnarray}}
\newcommand{\brr}{\begin{array}}
\newcommand{\err}{\end{array}}
\newcommand{\bc}{\begin{center}}
\newcommand{\ec}{\end{center}}
\newcommand{\nn}{\nonumber}

\newcommand{\alphas}{\alpha_{ s}}

\begin{document}
\setcounter{page}{1}
\begin{flushright}
LPTENS 93/28 \\
ROME 93/958 \\
ULB-TH 93/09\\
July 1993\\
\end{flushright}
\centerline{\bf{Scheme Independence of the Effective Hamiltonian}}
\centerline{\bf{for $b \rightarrow s \, \gamma$ and $b \rightarrow s \, g$
Decays}}
\vskip 1cm
\centerline{\bf{ M. Ciuchini$^{a,b}$,
E. Franco$^b$, G. Martinelli$^{b,c}$}}
\centerline{\bf{L. Reina$^d$ and L. Silvestrini$^b$ }}
\centerline{$^a$ INFN, Sezione Sanit\`a, V.le Regina Elena 299,
00161 Roma, Italy. }
\centerline{$^b$ Dip. di Fisica,
Universit\`a degli Studi di Roma ``La Sapienza" and}
\centerline{INFN, Sezione di Roma, P.le A. Moro 2, 00185 Roma, Italy. }
\centerline{$^c$ Laboratoire de Physique Th\'eorique de l'Ecole
Normale Sup\'erieure\footnote
{Unit\'e Propre du Centre National de
la Recherche Scientifique,
 associ\'ee \`a  l'\'Ecole Normale Sup\'erieure  et \`a
l'Universit\'e de Paris-Sud.}}
\centerline{24 rue Lhomond, 75231 Paris CEDEX 05, France.}
\centerline{$^d$ Service de Physique Th\'eorique\footnote{Chercheur IISN},
Universit\'e Libre de Bruxelles,}
\centerline{Boulevard du Triomphe, CP 225 B-1050 Brussels, Belgium.}
\begin{abstract}
We present a calculation of the
 effective weak Hamiltonian which governs $b \rightarrow s\, \gamma$
and $b \rightarrow s \, g$ transitions in two different renormalization
schemes  (NDR and HV). In the leading logarithmic approximation,
we show that the coefficients of the
effective Hamiltonian are scheme independent only when one takes correctly
into account the scheme dependence of one- and two- loop diagrams.
We demonstrate
that in NDR there are contributions which were missed in previous
calculations. These
contributions are necessary to obtain scheme independent coefficients
in the final results.
 \end{abstract}
\date{}
\newpage
\section{Introduction}
\label{sec:intro}
Radiative decays of B mesons have been the subject of several investigations
in the last years \cite{gsw1}-\cite{mi2}.
With the exception of two recent papers by M. Misiak\cite{mi1,mi2} and
\cite{yao}
 however, none of these studies has taken into account the
full set of leading logarithmic corrections. Moreover calculations
made in the dimensional reduction scheme (DRED)\cite{gri1,gri3,gri2}
seem to differ from the results obtained in the naive dimensional
regularization scheme (NDR)\cite{gsw1,ccrv,gsw2,mi1,mi2}.

In refs.\cite{ccrv} and \cite{mi1}, the calculation of refs.\cite{gsw1,gsw2}
was repeated in NDR and found in agreement with the
original calculation. However the origin of the difference between
the results obtained in NDR and DRED has never been clarified.
In the literature it is always implicitely (or explicitely) assumed
that the anomalous dimension matrix which enters in $b$ radiative decays
is regularization scheme independent, as usually happens in leading
logarithmic calculations. Thus, bearing in mind possible errors in the
calculation, the difference between NDR and DRED has remained so far
a mistery.

In this paper we have computed the anomalous dimension matrix relevant
for $b \rightarrow s \, \gamma$ and $b \rightarrow s \, g$
decays in two different regularization
schemes: NDR  and the t'Hooft-Veltman scheme (HV)\cite{hv}.
We show that the values of some diagrams are indeed regularization scheme
dependent. We explain the origin of the scheme dependence and show that the
regularization scheme independent effective Hamiltonian is obtained only by
taking into account the scheme dependence of some one loop diagrams, which
appear at zero order in $\alphas$. The contributions of these diagrams
were included in the calculation of Misiak\cite{mi1,mi2}, but their
role was not fully understood.
We show that
the easiest way to obtain the scheme independent result is to use the
HV regularization scheme. We also briefly discuss the role
of ``effervescent" operators and of
operators which vanish by the equations of motion. Finally we add some
comments on the gauge invariance  of the final result.

\section {Evolution and Initial Conditions for the Coefficients}
\label{sec:evo}
The effective Hamiltonian for $b \rightarrow s \, \gamma, \, g$ decays
can be written as:
\be
 H_{eff}= V_{tb} V^*_{ts} \frac {G_F} {\sqrt{2}}
\sum_{i=1}^{8} Q_{ i}(\mu) C_{ i}(\mu) \sim  \vec Q^T(\mu) \, \vec C(\mu)
\label{eh} \ee
where   $V_{ij}$ are
 the elements of the CKM\cite{cab,km} mixing matrix.
The operator basis is given by:
\bea
Q_{ 1}&=&({\bar s}_{\alpha}c_{\beta})_{ (V-A)}
    ({\bar c}_{\beta}b_{\alpha})_{ (V-A)}
   \nn\\
Q_{ 2}&=&({\bar s}_{\alpha}c_{\alpha})_{ (V-A)}
    ({\bar c}_{\beta}b_{\beta})_{ (V-A)}
\nn \\
Q_{ 3,5} &=& ({\bar s}_{\alpha}b_{\alpha})_{ (V-A)}
    \sum_{q=u,d,s,\cdots}({\bar q}_{\beta}q_{\beta})_{ (V\mp A)}
\nn \\
Q_{ 4,6} &=& ({\bar s}_{\alpha}b_{\beta})_{ (V-A)}
    \sum_{q=u,d,s,\cdots}({\bar q}_{\beta}q_{\alpha})_{ (V\mp A)}
\nn \\
Q_{ 7} &=& \frac{Q_de}{16\pi^2}m_b{\bar s}_{\alpha}\sigma^{\mu\nu}_{(V+A)}
           b_{\alpha}F_{\mu\nu}
\nn \\
Q_{ 8} &=& \frac{g}{16\pi^2}m_b{\bar s}_{\alpha}\sigma^{\mu\nu}_{(V+A)}
           t^A_{\alpha\beta}b_{\beta}G^A_{\mu\nu}
\label{basis}
\eea
In eq.(\ref{basis}) the subscript $(V\mp A)$ indicates the chiral structure,
$\alpha,\,\beta$ are colour indices, $m_b$ denotes the b quark mass, $Q_d=-
\frac{1}{3}$ is the electric charge of the down-type quarks and $g$
($e$) is
the strong (electro-magnetic) coupling. Our colour matrices are normalized in
such a way that $ Tr(t^A t^B)= \delta^{AB}/2$.

The coefficients $\vec C(\mu)$ obey the renormalization group equations:
\be
\left( - \frac {\partial} {\partial t} + \beta ( \alphas )
\frac {\partial} {\partial \alphas} - \frac {\hat \gamma^T ( \alphas ) }{2}
 \right) \vec C(t, \alphas(t)) =0 \label{rge} \ee
where $t=ln ( M_W^2 / \mu^2 )$ and $\alpha_s=g^2/4 \pi$.
The factor of $2$ in eq.(\ref{rge}) normalizes the anomalous dimension
matrix as in refs.\cite{bjlw1}-\cite{noi2}. $\hat \gamma^T$ includes
the contribution due to the renormalization of $m_b$ and, in the case of
$Q_8$, the contribution due to the renormalization of the gluon field
and of the strong coupling constant $g$, see e.g. ref.\cite{shif,ccrv}.

The anomalous dimension matrix  is ``almost'' triangular, in the
sense that the operators of dimension six can mix with the magnetic
operators $Q_7$ and $Q_8$, whereas  the dimension  five  operators $Q_{7,8}$
can only mix between themselves.
For later convenience we introduce a reduced
$6 \times 6$ anomalous dimension matrix, ${\hat\gamma}_r$, which mixes the
operators $Q_{1},Q_{2},\dots,Q_{6}$ among themselves and a 6-component column
vector
$\vec C_r(\mu)=\left( C_1(\mu),C_2(\mu),\dots,C_6(\mu) \right)$
on which ${\hat\gamma}_r$ acts. We also introduce two 6-component vectors,
related to the two-loop anomalous dimension matrix,
$\vec \beta_7=\left( \gamma_{17},\gamma_{27},\dots,\gamma_{67} \right)$
and
$\vec \beta_8=\left( \gamma_{18},\gamma_{28},\dots,\gamma_{68} \right)$.

In terms of these quantities the renormalization group
equations can be written as:
\bea
2\mu^2\frac {d}{d\mu^2}\vec C_r(\mu) &=& \frac{\alphas}{4\pi}
   \hat\gamma^T_r\vec C_r(\mu)
\nn \\
2\mu^2\frac {d}{d\mu^2} C_7(\mu) &=& \frac{\alphas}{4\pi}\left(
   \vec\beta_7\cdot\vec C_r(\mu)+\gamma_{77}C_7(\mu)+\gamma_{87}C_8(\mu)\right)
\nn \\
2\mu^2\frac {d}{d\mu^2} C_8(\mu) &=& \frac{\alphas}{4\pi}\left(
   \vec\beta_8\cdot\vec C_r(\mu)+\gamma_{88}C_8(\mu)\right)
\nn \\
\label{evolu1}
\eea
where $\alphas=\alphas(\mu)$ and $\mu^2 {d}/{d\mu^2}=
\mu^2 {\partial}/{\partial\mu^2}+\beta (\alpha_s){\partial}/{\partial
\alpha_s}$.
One can easily diagonalize the sub-matrix of magnetic operators and write the
renormalization group equations as:
\bea
2\mu^2\frac {d}{d\mu^2}\vec C_r(\mu) &=& \frac{\alphas}{4\pi}
   \hat\gamma^T_r\vec C_r(\mu)
\nn \\
2\mu^2\frac {d}{d\mu^2} v_7(\mu) &=& \frac{\alphas}{4\pi}\gamma_{77}v_7(\mu)
\nn \\
2\mu^2\frac {d}{d\mu^2} v_8(\mu) &=& \frac{\alphas}{4\pi}\gamma_{88}v_8(\mu)
\label{evolu2}
\eea
where:
\bea
v_7(\mu) &=& C_7(\mu)+\vec\alpha_7\cdot\vec C_r(\mu)+\frac{\gamma_{87}}
 {\gamma_{77}-\gamma_{88}}C_8(\mu)
\nn\\
v_8(\mu) &=& C_8(\mu)+\vec\alpha_8\cdot\vec C_r(\mu)
\label{eigenvec}
\eea
with
\bea
\vec\alpha_7 &=& \left(\gamma_{77}\hat 1-\hat\gamma_r\right)^{-1}\left[
  \vec\beta_7+\frac{\gamma_{87}}{\gamma_{77}-\gamma_{88}}\vec\beta_8\right]
\nn\\
\vec\alpha_8 &=& \left(\gamma_{88}\hat 1-\hat\gamma_r\right)^{-1}
   \vec\beta_8
\label{alphas}
\eea

Contrary to the claims of refs. \cite{gsw1}-\cite{ccrv},\cite{mi1},
$\vec\beta_7$ and $\vec\beta_8$ do depend on the regularization, as our
calculations in NDR and HV  explicitly show, see below. A
regularization dependent anomalous dimension matrix in a leading logarithmic
(but two loop) calculation is an exotic phenomenon and deserves some
explanations. At one loop, but at order $\alphas^0$, the operators $Q_5$
and $Q_6$ can mix with $Q_7$ and $Q_8$ with finite coefficients, through
the diagrams shown in fig.\ref{fig:oneloop}. In the figure, the  cross
denotes a mass ($m_b$) insertion in the heavy quark propagator of the
penguin loop.
The coefficients of the mixing are regularization dependent. They vanish in
HV (and DRED), but not in NDR. As a consequence, the values of all two loop
diagrams contributing to $\vec\beta_7$ and $\vec\beta_8$, which have as
sub-diagrams those in fig.\ref{fig:oneloop}, will depend on the
regularization. Furthermore the diagrams in fig.\ref{fig:massloop}, which
vanish in HV, have instead to be considered in NDR.
%___________________________________________________________________________
\begin{figure}[t]   % produce figure here
    \begin{center}
       \setlength{\unitlength}{1truecm}
       \begin{picture}(5.0,4.0)
          \put(-6.5,-20.5){\special{oneloop.ps}}
       \end{picture}
    \end{center}
    \caption[]{One loop diagrams which mix $Q_5$ and $Q_6$ with $Q_7$
and $Q_8$. The mixing appears at order $\alphas^0$ and the coefficients
of the mixing are regularization scheme dependent.}
    \protect\label{fig:oneloop}
\end{figure}
%___________________________________________________________________________

 An example of a diagram which depends on the regularization
 is reported in fig.\ref{fig:d13ex}.
The integration over the penguin loop, where a bare four fermion
operator is inserted,
will transform it in an effective $\bar s \gamma^{\mu}_L D^\nu  t^A
G^A_{\mu\nu}
 b$ local vertex ($\gamma^{\mu}_{R,L}=\gamma^{\mu} ( 1
\pm \gamma_5)$). This effective vertex is, by the equations of motion,
equivalent to $\bar s\gamma^{\mu}_L t^A b \sum_q \frac{1}{2}\left(\bar q
\gamma_{\mu L} t^A q+\bar q\gamma_{\mu R} t^A q\right)$, which can
be written as a linear
combination of $Q_{3},\dots,Q_{6}$. Thus, the divergent part of the coefficient
of
the effective vertex combines with the finite, regularization dependent,
mixing coefficient of $Q_{5,6}$ with $Q_{7,8}$. This means that it will give
a scheme dependent contribution to $\vec\beta_7$ and $\vec\beta_8$.
%___________________________________________________________________________
\begin{figure}[t]   % produce figure here
    \begin{center}
       \setlength{\unitlength}{1truecm}
       \begin{picture}(6.0,7.0)
          \put(-7.5,-15.5){\special{d13ex.ps}}
       \end{picture}
    \end{center}
    \caption[]{Two loop diagram contributing to the mixing of the operators
$Q_{1,6}$ with $Q_7$ and $Q_8$. The value of the diagram is regularization
scheme dependent.
This figure explains the scheme dependence of the diagram.
 The integration over the penguin loop mixes the
original operator $Q_i$ with $\bar s \gamma^{\mu}_L D^\nu G^A_{\mu\nu}
t^A b$, denoted by a big square. By the equations of motion, this
operator is equivalent to a four fermion operator, denoted by a small
square in the figure, see also the text.}
    \protect\label{fig:d13ex}
\end{figure}
%___________________________________________________________________________

We now show how, from the finite one loop coefficients and the two loop
anomalous dimension matrix, which are separately scheme dependent, a physical,
scheme independent, effective Hamiltonian is obtained.
Let us consider the finite one loop mixing of the operators $Q_5$ and
$Q_6$ with $Q_7$:
\bea
<s\gamma|H_{eff}|b>&=&C_7(\mu)<s\gamma|Q_7(\mu)|b>+
C_5(\mu) <s\gamma|Q_5(\mu)|b> \nn\\
&&+ C_6(\mu) <s\gamma|Q_6(\mu)|b>
\eea
We can interpret the non vanishing of the one loop matrix elements of $Q_5$
and $Q_6$ as the effect of a mixing matrix acting at order $\alphas^0$,
see below:
\be
<s\gamma|H_{eff}|b>=\tilde C_7(\mu)<s\gamma|Q_7(\mu)|b>
\label{matele1}
\ee
where $\tilde C_7(\mu)=C_7(\mu)+\vec Z_7\cdot\vec C_r(\mu)$.
The vector $\vec Z_7$ (and $\vec Z_8$, see below)
is regularization dependent.
$\vec Z_7$
vanishes in HV (and DRED) and
\be \vec Z_7\equiv \left(0,0,0,0,2,2N\right) \label{zv7} \ee
in NDR. Eq.(\ref{matele1}) corresponds to a finite renormalization of the
operators such that the matrix elements of $Q_5$ and $Q_6$ vanish.
In this way the renormalized operators are the same in NDR and HV.
Similarly one finds:
\bea
<sg|H_{eff}|b>&=&C_8(\mu)<sg|Q_8(\mu)|b>+
C_5(\mu) <sg|Q_5(\mu)|b> \nn\\
&&=\tilde C_8(\mu)<sg|Q_8(\mu)|b>
\label{matele2}
\eea
with $\tilde C_8(\mu)=C_8(\mu)+\vec Z_8\cdot\vec C_r(\mu)$. $\vec Z_8=\vec 0$
in HV (and DRED) and
\be\vec Z_8\equiv \left(0,0,0,0,2,0\right) \label{zv8} \ee in NDR.

The renormalization group equations (\ref{evolu2}) can be easily solved
and combined with eqs.(\ref{matele1}) and (\ref{matele2})
to obtain the effective Hamiltonian expressed in terms of operators
renormalized at the scale $\mu$:
\bea
H_{eff}&\sim&\vec Q_r^T(\mu)\cdot \vec C_r(\mu) \nn \\
&&+\left\{ v_7(\mu)+\left[\left( \hat\gamma_r-\gamma_{77}\hat 1\right)^{-1}
\left(\vec\beta_7+\frac{\gamma_{87}}{\gamma_{77}-\gamma_{88}}\vec\beta_8
\right)+\vec Z_7\right]\cdot\vec C_r(\mu) \right. \nn \\
&&+\left. \frac{\gamma_{87}}{\gamma_{88}-\gamma_{77}}\left[v_8(\mu)+
\left(\hat\gamma_r-\gamma_{88}\hat 1\right)^{-1}\vec\beta_8\cdot
\vec C_r(\mu)\right]\right\}Q_7(\mu) \nn \\
&&+\left\{v_8(\mu)+\left[
\left(\hat\gamma_r-\gamma_{88}\hat 1\right)^{-1}\vec\beta_8 + \vec Z_8
\right]\cdot\vec C_r(\mu)\right\}Q_8(\mu) \nn \\
\\
&=&\vec Q_r^T(\mu)\cdot \vec C_r(\mu) \nn \\
&&+\left\{ v_7(\mu)+\frac{\gamma_{87}}{\gamma_{88}-\gamma_{77}}v_8(\mu)+
\left[\left( \hat\gamma_r-\gamma_{77}\hat 1\right)^{-1}
\left(\vec\beta_7+\frac{\gamma_{87}}{\gamma_{77}-\gamma_{88}}\vec\beta_8
\right)+\vec Z_7\right.\right. \nn \\
&&\left.\left.+\frac{\gamma_{87}}{\gamma_{77}-\gamma_{88}}\vec Z_8
\right]\cdot\vec C_r(\mu)+
\frac{\gamma_{87}}{\gamma_{88}-\gamma_{77}}\left[
\left(\hat\gamma_r-\gamma_{88}\hat 1\right)^{-1}\vec\beta_8+\vec Z_8\right]
\cdot\vec C_r(\mu)\right\}Q_7(\mu) \nn \\
&&+\left\{v_8(\mu)+\left[
\left(\hat\gamma_r-\gamma_{88}\hat 1\right)^{-1}\vec\beta_8 + \vec Z_8
\right]\cdot\vec C_r(\mu)\right\}Q_8(\mu)
\label{heff}
\eea

$\vec C_r(\mu)$, $v_7(\mu)$ and $v_8(\mu)$ are scheme independent quantities
because they obey the scheme independent renormalization group equations
(\ref{evolu2}).
$\vec\beta_{7,8}$ and $\vec Z_{7,8}$ are scheme dependent quantities.
We now prove that the combinations
\bea
\vec\omega_7&=&\left( \hat\gamma_r-\gamma_{77}\hat 1\right)^{-1}
\left(\vec\beta_7+\frac{\gamma_{87}}{\gamma_{77}-\gamma_{88}}\vec\beta_8\right)
+\vec Z_7+\frac{\gamma_{87}}{\gamma_{77}-\gamma_{88}}\vec Z_8 \nn \\
\vec\omega_8&=&\left(\hat\gamma_r-\gamma_{88}\hat 1\right)^{-1}\vec\beta_8
+ \vec Z_8
\label{schind}
\eea
appearing in eq.(\ref{heff}) are instead scheme independent.
Let us introduce the renormalization matrix $\hat Z^a$ which gives
the renormalized operators $\vec Q(\mu)$ in terms of the bare operators
$\vec Q_B$, in the regularization labelled as ``$a$'' :
\be
\vec  Q(\mu)=\left(\hat Z^a\right)^{-1}\vec Q_B
\ee
The anomalous dimension matrix is defined in terms of $\hat Z^a$ as:
\be
\hat\gamma^a=2\left(\hat Z^a(\mu)\right)^{-1}\mu^2\frac{d}{d\mu^2}\hat Z^a(\mu)
\label{diman}
\ee

If one uses two different regularization schemes, ``$a$'' and ``$b$'' say,
the renormalization constants are related through the equation:
\be
\hat Z^a=\hat Z^b\hat r
\label{renmat}
\ee
from eqs.(\ref{diman}) and (\ref{renmat}) we obtain:
\be
\hat\gamma^a=\hat r^{-1}\hat\gamma^b\hat r
\label{check}
\ee
In other cases, as for example for the operators appearing in the
$\Delta S=1$ weak Hamiltonian, $\hat r$ differs from
the identity matrix by terms of order $\alphas$, i.e.
$\hat r=\hat 1+frac{\alphas}{4\pi}\Delta r$ \cite{bjlw1}-\cite{acmp}.
In the present calculation instead, because of the diagrams in
fig.\ref{fig:oneloop}, $\hat r$ differs from $\hat 1$ by terms
of order $\alphas^0$, see eqs.(\ref{matele1}) and (\ref{matele2}).
The matrix $\hat r$ can be written as:
\be \hat r=\left(\begin{array}{c c }
\hat 1_6 &  -\Delta \hat Z \\
0 & \hat 1_2
\end{array}\right)
\ee
where
$\hat 1_{6,2}$ are $6\times 6$ and $2\times 2$ identity matrices and
$\Delta\hat Z$ is a $6\times 2$ matrix. Its two columns are given by the
difference of the vectors $\vec Z_7$ and $\vec Z_8$ in the regularizations
``$a$'' and ``$b$'', $ \Delta \hat Z= (\vec Z_7^a-\vec Z_7^b,
\vec Z_8^a-\vec Z_8^b)$.
At the order we are interested in,  eq.(\ref{check})  gives:
\be
\hat\gamma^a-\hat\gamma^b=[\Delta\hat Z,\hat\gamma^b]-\Delta\hat Z\hat\gamma^b
\Delta\hat Z
\label{check1}
\ee
Eq.(\ref{check1}), which connects the anomalous dimension matrices
in two different regularizations, holds diagram by diagram and
can be used as a check of the calculation of any single diagram in
different regularization schemes \cite{noi2}. More details will be given
in ref.\cite{noibis}.

Taking into account the structure of the matrix $\Delta\hat Z$ we find:
\bea
\hat\gamma_r^a&=&\hat\gamma^b_r=\hat\gamma_r\nn\\
(\Delta \vec \beta_7)_j=
\Delta\gamma_{j7}&=&\left[\left(\gamma_{77}\hat 1-\hat\gamma_r\right)\Delta\vec
Z_7+\gamma_{87}\Delta\vec Z_8 \right]_j\nn \\
(\Delta \vec \beta_8)_j=
\Delta\gamma_{j8}&=&\left[\left(\gamma_{88}\hat 1-\hat\gamma_r\right)
\Delta\vec Z_8\right]_j
\label{check2}
\eea
where $j$ denotes the components of the vectors $\Delta \vec \beta_{7,8}$,
$j=1, \dots , 6$.
Eq.(\ref{check2}) demonstrates that the combinations $\vec\omega_7$ and $\vec
\omega_8$ appearing in eq.(\ref{schind}) are regularization scheme independent.
Since  $\vec Z_7$ and $\vec Z_8$
are zero in HV, in the present case $\Delta  \hat Z$
is  simply given by $\vec Z_{7,8}$
computed in NDR ($\Delta\hat\gamma=\hat\gamma^{NDR}-\hat\gamma^{HV}$
 and $\Delta\hat Z=
\hat Z^{NDR}-\hat Z^{HV}=\hat Z^{NDR}$). We have computed the anomalous
dimension in both NDR and HV \footnote
{The reduced mixing matrix $\hat\gamma_r$ has been recently computed up to
the next-to-leading order in refs.\cite{bjlw1}-\cite{noi2}.}.
{}From the one-loop diagrams in
fig.\ref{fig:oneloop} and the two-loop diagrams in fig.\ref{fig:diag1}
and \ref{fig:massloop},  we obtain the anomalous dimension matrices
(\ref{mathv}), (\ref{matndr}), given below. Combining (\ref{mathv})
and (\ref{matndr})
with eqs.(\ref{zv7},\ref{zv8}), one indeed finds the same $\vec \omega_{7,8}$
in the NDR and HV schemes.

In refs.\cite{gsw1,ccrv,gsw2}, they considered a restricted operator basis, in
the NDR scheme.  Our results for the two loop Feynman diagrams
agree with those reported there.
 The anomalous dimension matrix given
in these references is  however
regularization scheme dependent. The scheme dependence
follows from the non-vanishing of the diagrams in fig.\ref{fig:oneloop}, which
spoil the equations of motion of the regularized theory:
\be
\bar s \gamma^{\mu}_L D^\nu  t^A G^A_{\mu\nu}
 b =\bar s\gamma^{\mu}_L t^A b \sum_q \frac{1}{2}\left(\bar q
\gamma_{\mu L} t^A q+\bar q\gamma_{\mu R} t^A q\right).
\ee
Indeed, using the NDR scheme,
 the values of the two diagrams
in fig.\ref{fig:mismatch} differ, because of the contributions coming from
the insertion of $m_b$ in the loop propagators,
see also fig.\ref{fig:oneloop}.
It is easy to show that the scheme dependence can be eliminated
 by enforcing  the equations of motion with a suitable, finite
subtraction, at order $\alpha_s^0$.
%___________________________________________________________________________
\begin{figure}[t]   % produce figure here
    \begin{center}
       \setlength{\unitlength}{1truecm}
       \begin{picture}(6.0,5.5)
          \put(-7,-18.0){\special{mismatch.ps}}
       \end{picture}
    \end{center}
    \caption[]{Diagrams in the reduced $(a)$ and complete $(b)$ bases.
In $(a)$ we insert the operator $\bar s \gamma^{\mu}_L D^\nu  t^A G^A_{\mu\nu}
 b$,
in $(b)$ the corresponding four fermion
operators. Because
of  mass insertions, which spoil the equations of motion, in NDR the two
diagrams have different values.  On the contrary, they give the same result
in the HV scheme.}
    \protect\label{fig:mismatch}
\end{figure}
%___________________________________________________________________________

In NDR, when one uses the complete basis (\ref{basis}), the  diagrams  in
fig.\ref{fig:massloop} have to be computed\cite{mi1}.
These diagrams vanish in the HV scheme.
In NDR instead they give a non-zero contribution
to the anomalous dimension, precisely for the same reason that the diagrams in
fig.\ref{fig:oneloop} do not vanish. In the final answer (\ref{heff}),
the contribution of each of the
diagrams in fig.\ref{fig:massloop}
is exactly cancelled by the corresponding term in $\vec Z_7$ and
$\vec Z_8$, as it can be checked diagram by diagram. In fact the diagrams in
fig.\ref{fig:massloop} give a contribution which goes like $-\left(\hat
\gamma_r^T-\gamma_{77,88}\hat 1\right)\vec Z_{7,8}$. The only exceptions
are the diagrams $P_4$ and $F_4$
in fig.\ref{fig:diag1}\footnote{This diagram would
be present also in the reduced operator basis used in
refs.\cite{gsw1,ccrv,gsw2}.}
 (see also fig.\ref{fig:d13ex}). In this case,
 $\vec Z_{7,8}$ cancel the regularization
dependent part of that two-loop diagram, giving the same result one would
obtain in HV.
In HV, since $\vec Z_7$ and $\vec Z_8$
vanish, one obtains the regularization scheme independent result directly
from the evaluation of the diagrams in fig.\ref{fig:diag1}.
Those calculations,
\cite{gsw1,ccrv} which did not  include the effect of
$\vec Z_{7,8}$ in the effective Hamiltonian, are not correct because
 of the regularization dependence of the diagrams $P_4$ and $F_4$
in fig.\ref{fig:diag1}.
On the other hand, for the complete basis, it is necessary to include
the contribution of the diagrams in fig.\ref{fig:massloop}.

%___________________________________________________________________________
\begin{figure}[t]   % produce figure here
    \begin{center}
       \setlength{\unitlength}{1truecm}
       \begin{picture}(6.0,9.0)
          \put(-7.5,-12.5){\special{diag1.ps}}
       \end{picture}
    \end{center}
    \caption[]{Diagrams which contribute both in the NDR and HV schemes.
In the evaluation of the diagrams counter-terms and ``effervescent"  operators
have been included, see also ref.\cite{noibis}.}
    \protect\label{fig:diag1}
\end{figure}
%___________________________________________________________________________

We now summarize our discussion. In HV, where $\vec
Z_{7,8}$ vanish, we have only to compute the contribution to the anomalous
dimension matrix of the diagrams in fig.\ref{fig:diag1} (besides the usual
one loop diagrams which mix the operators $Q_{1},\dots,Q_{6}$). The resulting
anomalous
dimension matrix is regularization independent and does not require any further
manipulation. In NDR, the two loop anomalous dimension has to be computed from
the diagrams in figs.\ref{fig:diag1} and \ref{fig:massloop}. The result
is then combined with $\vec Z_{7,8}$ as in eq.(\ref{heff}). The final result
is then identical to the result obtained in HV,
as we have explicitely checked. This implies that our results for the anomalous
dimension matrices in NDR and HV, (\ref{mathv})-(\ref{matndr}),
satisfy the general relation (\ref{check1}).
We report below the anomalous dimension matrix in HV,
to be used for the evolution of the coefficients,
together with the initial conditions\cite{desh}
\bea
C_2(M_W)&=&1 \nn\\
C_{1},C_{3},\dots,C_{6}(M_W)&=&0 \nn\\
C_{ 7} (M_{ W}) &=&  -3\frac{3
x^{ 3} - 2 x^{ 2}}{2(1 -
x)^{ 4}}\ln {x} - \frac{8 x^{ 3}+ 5
x^{ 2} - 7 x}{4 (1 - x)^{ 3}} \\
C_{ 8} (M_{ W}) &=&
-\frac{3 x^{ 2}}{2 (1-x)^{
4}}\ln {x} + \frac{x^{ 3}-5 x^{
2}- 2 x}{4 (1 - x )^{ 3}} \nn
\label{inicoef}
\eea
where $x=m_t^2/M_W^2$.
\begin{eqnarray}
\hat{\gamma}_{\scriptscriptstyle ai}&=&\frac{\alphas}{4\pi}
\left( \begin{array}{cccccc}
-\frac{6}{N} & 6 &0 &0 &0 &0 \\
 & & & & & \\
6 &-\frac{6}{N} & -\frac{2}{3N} & \frac{2}{3} & -\frac{2}{3N} & \frac{2}{3} \\
 & & & & & \\
0 &0 &-\frac{22}{3N} &\frac{22}{3} &-\frac{4}{3N} &\frac{4}{3} \\
 & & & & & \\
0 &0 &6-\frac{2n_{\scriptscriptstyle f}}{3N}&-\frac{6}{N} + \frac{2
n_{\scriptscriptstyle f}}{3}&-\frac{2n_{\scriptscriptstyle f}}{3N} &\frac{2
n_{\scriptscriptstyle f}}{3}\\
 & & & & & \\
0 & 0 & 0 &0 &\frac{6}{N} &-6 \\
 & & & & & \\
0 &0 &-\frac{2n_{\scriptscriptstyle f}}{3N} & \frac{2n_{\scriptscriptstyle
f}} {3}  &-\frac{2n_{\scriptscriptstyle f}}{3N} &
-12\frac{N^{\scriptscriptstyle 2} -1}{2N} + \frac{2n_{\scriptscriptstyle
f}}{3} \\
 & & & & & \\
0 &0 &0 &0 &0 &0 \\
 & & & & & \\
0 &0 &0 &0 &0 &0 \\
\end{array} \right) \nonumber \\
 & & \nonumber \\
 & & \nonumber \\
\hat{\gamma}^{\scriptscriptstyle HV}_{\scriptscriptstyle a \alpha}&=
&\frac{\alphas}{4\pi}
\left( \begin{array}{cc}
0 & 6   \\
 & \\
\frac{8}{9} \frac{N^{\scriptscriptstyle 2}-1}{2N} +
\frac{12Q_{\scriptscriptstyle
u}}{Q_{\scriptscriptstyle d}}\frac{N^{\scriptscriptstyle 2}-1}{2N} &
\frac{22N}{9} - \frac{58}{9N}   \\
 & \\
 \frac{232}{9}
\frac{N^{\scriptscriptstyle 2}-1}{2N} & \frac{44N}{9} - \frac{116}{9N} + 6 n
_{\scriptscriptstyle f}   \\
 & \\
\frac{8n_{\scriptscriptstyle f}}{9}\frac{N^{\scriptscriptstyle 2}-1}{2N}
+  \frac{12\bar{n}_{\scriptscriptstyle f}(N^{\scriptscriptstyle 2}-1)}{2N}
 & 12 + \frac{22Nn_{\scriptscriptstyle f}}{9}
- \frac{58n_{\scriptscriptstyle f}}{9N}\\
 & \\
-16\frac{N^{\scriptscriptstyle 2}-1}{2N} &
-4N + \frac{8}{N} - 6 n_{\scriptscriptstyle f}   \\
 & \\
\frac{8n_{\scriptscriptstyle f}
}{9}\frac{N^{\scriptscriptstyle 2}-1}{2N}
 - \frac{12\bar{n}_{\scriptscriptstyle f}(N^{\scriptscriptstyle 2}-1)}{2N}
 & -8 -\frac{32Nn_{\scriptscriptstyle f}}{9}
+ \frac{50n_{\scriptscriptstyle f}}{9N}  \\
 & \\
8\frac{N^{\scriptscriptstyle 2}-1}{2N} & 0  \\
 & \\
8\frac{N^{\scriptscriptstyle 2}-1}{2N}  & 4N -\frac{8}{N}   \\
\end{array} \right)
\label{mathv}
\end{eqnarray}
where $n_f=n_u+n_d$ is the number of active flavours,
$N$ is the number of colours,
$a=1,\dots,8$, $i=1,\dots,6$ and $\alpha=7,8$.
In the NDR scheme we obtain
$\hat{\gamma}^{\scriptscriptstyle NDR}_{\scriptscriptstyle ai} =
\hat{\gamma}^{\scriptscriptstyle HV}_{\scriptscriptstyle ai}$ and:
\bea
\hat{\gamma}^{\scriptscriptstyle NDR}_{\scriptscriptstyle
a \alpha}&=&\frac{\alphas}{4\pi}
\left( \begin{array}{cc}
0 & 6   \\
 & \\
\frac{-16}{9} \frac{N^{\scriptscriptstyle 2}-1}{2N} +
\frac{12Q_{\scriptscriptstyle
u}}{Q_{\scriptscriptstyle d}}\frac{N^{\scriptscriptstyle 2}-1}{2N} &
\frac{22N}{9} - \frac{46}{9N}   \\
 & \\
 \frac{184}{9}
\frac{N^{\scriptscriptstyle 2}-1}{2N} & \frac{44N}{9} - \frac{92}{9N} + 6 n
_{\scriptscriptstyle f}   \\
 & \\
-\frac{16n_{\scriptscriptstyle f}}{9}\frac{N^{\scriptscriptstyle 2}-1}{2N}
+  \frac{12\bar{n}_{\scriptscriptstyle f}(N^{\scriptscriptstyle 2}-1)}{2N}
 & 12 + \frac{22Nn_{\scriptscriptstyle f}}{9}
- \frac{46n_{\scriptscriptstyle f}}{9N}\\
 & \\
40\frac{N^{\scriptscriptstyle 2}-1}{2N} &
4N - \frac{20}{N} - 6 n_{\scriptscriptstyle f}   \\
 & \\
-\frac{16n_{\scriptscriptstyle f}
}{9}\frac{N^{\scriptscriptstyle 2}-1}{2N}
 - \frac{12\bar{n}_{\scriptscriptstyle f}(N^{\scriptscriptstyle 2}-1)}{2N}
+\frac{40N(N^{\scriptscriptstyle 2}-1)}{2N}
 & -8 -\frac{32Nn_{\scriptscriptstyle f}}{9}
+ \frac{62n_{\scriptscriptstyle f}}{9N}  \\
 & \\
8\frac{N^{\scriptscriptstyle 2}-1}{2N} & 0  \\
 & \\
8\frac{N^{\scriptscriptstyle 2}-1}{2N}  & 4N -\frac{8}{N}   \\
\end{array} \right) \nonumber
\label{matndr}
\eea
where $\bar n_f=n_d+\frac{Q_u}{Q_d}n_u$.

Let us compare our results with previous calculations.

In refs.
\cite{gsw1}-\cite{gsw2} it was assumed
that the values of the two loop diagrams were scheme independent. In the
NDR scheme, we agree with the results of refs.\cite{gsw1,ccrv,gsw2} when
we reduce to the basis used in those references. However
the necessity of adding $\vec Z_{7,8}$, to obtain a scheme
independent result, was missed.
%___________________________________________________________________________
\begin{figure}[t]   % produce figure here
    \begin{center}
       \setlength{\unitlength}{1truecm}
       \begin{picture}(6.0,9.0)
          \put(-7.5,-12.5){\special{massloop.ps}}
       \end{picture}
    \end{center}
    \caption[]{Diagrams which have to be considered only in the NDR scheme.
In the evaluation of the diagrams counter-terms and ``effervescent"  operators
have been included.}
    \protect\label{fig:massloop}
\end{figure}
%___________________________________________________________________________

In ref.\cite{mi1}
Misiak realized the necessity of adding $\vec Z_7$, even if he did not
understand its role for the scheme independence of the final
result.
Moreover the paper contains several errors:
counter-terms of current-current diagrams and
``effervescent" operators\footnote{Without effervescent operators
relation (\ref{check1}) between HV and NDR would fail.}
were not taken into account
and there are inconsistencies
between different anomalous matrix elements, which can be
related to the same set of diagrams. In ref.\cite{mi2},
some of these errors have been corrected: counter-terms of current-current
diagrams and
effervescent operators have been considered and included in the final result.
However, our results (\ref{matndr}) and the matrix given in eq.(21)
of ref.\cite{mi2} do not agree. By looking to colour factors,
the disagreement seems to originate from
diagrams $P_2$ and $P_3$ in fig.\ref{fig:diag1}
with a mass ($m_b$) insertion in
the loop propagators.

The calculation done in DRED in ref.\cite{gri1,gri3,gri2}
could have lead in principle to the correct results.
The authors however were biased by the idea that the two loop calculation
had to be regularization scheme independent and interpreted the difference
between NDR and DRED as a failure of the DRED scheme. We have demonstrated
that there are not priviliged schemes and that all the
schemes, if correctly used, give the same results for the Wilson
coefficients\footnote{At this order in the perturbative expansion.}.
We are actually performing the calculation also in the DRED scheme to show
that this is indeed the case.

We were unable to understand how the calculation of ref.\cite{yao}
was done. We disagree with the final result for the anomalous dimension
matrix given in this reference. The results of ref.\cite{yao} differ also
from the results of ref.\cite{mi2}.

 We want  to
add a few comments on the operator basis and  gauge invariance.
In refs.\cite{gri1} and \cite{ccrv} a redundant basis of operators which
vanish (or coincide) by the equations of motion was used. In ref.
\cite{noi2},
it was shown that this leads to an unnecessary complication and that the
correct result can be obtained by working directly with a reduced basis
of indipendent operators, as the basis in eq.(\ref{basis}). This remains
true in the present calculation, as we have explicitly checked.
As noticed
already in ref.\cite{mi1,mi2}, it is not necessary to work in the background
field gauge  to obtain a gauge invariant result. We have done (for the
non-abelian diagrams) the calculation in the Feynman and background gauges
(see also ref.\cite{noi2}) with identical final results. We plan to present
a phenomenological analysis in a separate
publication, together with all the details of the calculation
(treatment of the effervescent operators, counter-terms, gauge
invariance, contribution of any single diagram, etc.)
and the results in the DRED scheme.

\section*{Acknowledgments}
M. C., E. F., L. R. and L. S. thank for the hospitality the LPT at ENS, Paris,
where part of this work has been done. We acknowledge the partial support of
the
MURST, Italy, and INFN.


\begin{thebibliography}{88}
\bibitem{gsw1}  B. Grinstein, R. Springer, M.B. Wise,
\pl{B202} (1988) 138.
\bibitem{gri1}  R. Grigjanis, P.J. O'Donnel, M. Sutherland
and H. Navelet, \pl {213} (1988) 355.
\bibitem{gri3}  R. Grigjanis, P.J. O'Donnel, M. Sutherland
and H. Navelet, \pl{B223} (1989) 239.
\bibitem{gri2}  R. Grigjanis, P.J. O'Donnel, M. Sutherland
and H. Navelet, \pl{B237} (1990) 252.
\bibitem{ccrv} G. Cella, G. Curci, G. Ricciardi and  A. Vicer\'e,
\pl{B248} (1990) 181.
\bibitem{gsw2} B. Grinstein, R. Springer, M.B. Wise,
\np{B339} (1990) 269.
\bibitem{crv} G. Cella, G.  Ricciardi, A. Vicer\'e,
\pl{B258} (1991) 212.
\bibitem{mi1} M. Misiak, \pl{B269} (1991) 161.
\bibitem{mi2} M. Misiak, \np{B393} (1993) 23.
\bibitem{yao} K. Adel and Y.P. Yao UM-TH-92/32 (1992).
\bibitem{hv} G. 't Hooft and M. Veltman,
\np {B44} (1972) 189.
\bibitem{cab} N. Cabibbo,
Phys. Rev. Lett. \underline{10} (1963) 1802.
\bibitem{km} J. Kobayashi and M. Maskawa,
Prog. Theor. Phys. \underline{49} (1973) 652.
\bibitem{bjlw1} A.J. Buras, M. Jamin, M.E. Lautenbacher and P.H. Weisz,
\np {B370} (1992) 69.
\bibitem{bjlw2} A.J. Buras, M. Jamin,  M.E. Lautenbacher and P.H. Weisz,
MPI-PAE/PTh 106/92 and TUM-T31-18/92.
\bibitem{bjl} A.J. Buras, M. Jamin and M.E. Lautenbacher,
MPI-PAE/PTh 107/92 and TUM-T31-30/92.
\bibitem{noi2} M. Ciuchini, E. Franco, G. Martinelli and L. Reina,
LPTENS 93/11, ROME prep. 92/913 and ULB-TH 93/03.
\bibitem{shif} M.A. Shifman, A.I. Vainshtein and V.I. Zakharov,
\prd {18} (1978) 2583.
\bibitem{acmp} G. Altarelli, G. Curci, G. Martinelli, S. Petrarca,
\np{B187} (1981) 461.
\bibitem{desh} N.G. Deshpande and M. Nazerimonfared, \np {B213} (1983) 390.
\bibitem{noibis} M. Ciuchini, E. Franco, G. Martinelli, L. Reina
and L. Silvestrini, in preparation.
\end{thebibliography}
\end{document}